# Stress-controlled avoided-crossing evolution of an optical microresonator induced by inter-fibre side coupling

Isha Sharma* and Misha Sumetsky
*Aston Institute of Photonic Technologies, Aston University, Birmingham B4 7ET, UK*
*i.sharma1@aston.ac.uk

## Abstract

Analogies between quantum mechanics and photonics provide a powerful means of transferring concepts and phenomena between the two fields. Photonic analogues offer a distinctive advantage by allowing us to engineer and continuously vary parameters fixed by the microscopic structure of quantum systems, revealing richer behaviour. Building on these analogies, the recent demonstration of widely tunable microresonators formed by side coupling between optical fibres marked a significant advance in the development of reconfigurable resonant photonic systems. However, those methods offered no direct way to reach the centre of the avoided crossing, where the uncoupled fibre cutoff frequencies coincide and the modes hybridise most strongly. Here, we solve this problem introducing an additional tuning degree of freedom by axially stretching a straight fibre coupled to a bent fibre. At essentially fixed coupling geometry, this lets us sweep the detuning between their uncoupled cutoff frequencies through the centre of their avoided crossing. As the avoided crossing is approached, the induced microresonator's resonance band widens. Beyond the crossing, it narrows and, strikingly, appears to 'hang' below the stretched fibre's rising cutoff frequency. Our experimental results agree with coupled-mode theory and indicate the feasibility of ∼100 MHz-scale control of the straight-fibre cutoff frequency and resonator spectral height over a ∼100 GHz tuning range using a relative elongation of order $10^{-3}$.

# Introduction

In many optical systems, the equations governing optical fields have the same mathematical form as those governing quantum wave functions, enabling photonic structures to emulate quantum phenomena in settings where their evolution can be directly observed and controlled [1, 2]. Optical analogues include tunnelling [3, 4], coherent state transfer [5], quantum Zeno dynamics [6], Bloch oscillations [7], avoided crossings and the associated Landau-Zener transitions [8–11], and Anderson [12] and dynamic localization [13].

Among these analogies, of particular interest is an avoided crossing between the eigenenergies of coupled quantum states, which, in photonics, occurs between the eigenfrequencies of coupled optical modes [8]. In atomic and molecular scattering, two electronic configurations $\psi^{(1)}_{\alpha_1}$ and $\psi^{(2)}_{\alpha_2}$ (Fig. 1a) may have *diabatic interaction potentials* $V_{11}(R)$ and $V_{22}(R)$ that vary with the separation $R$ and would cross if the configurations were uncoupled (Fig. 1b). An interaction between these states mixes them and removes the degeneracy, causing the corresponding *adiabatic potentials*, $U^{+}(R)$ and $U^{-}(R)$, to repel and form an avoided crossing – a fundamental mechanism in slow atomic and molecular collisions [14, 15]. Relative motion through the crossing region can induce nonadiabatic transitions between the electronic states, including Landau-Zener transitions [14, 15]. The interaction can also reshape the adiabatic potentials and create potential wells that support bound states. In pairs of Rydberg atoms, for example, multipolar coupling between attractive and repulsive pair states can create long-range potential wells supporting Rydberg macrodimers [16, 17]. Photonic Landau-Zener dynamics have similarly been proposed in curved directional couplers [9], predicted for interband transfer at a refractive-index step in waveguide arrays [10], and directly imaged in a femtosecond-laser-written coupler [11].

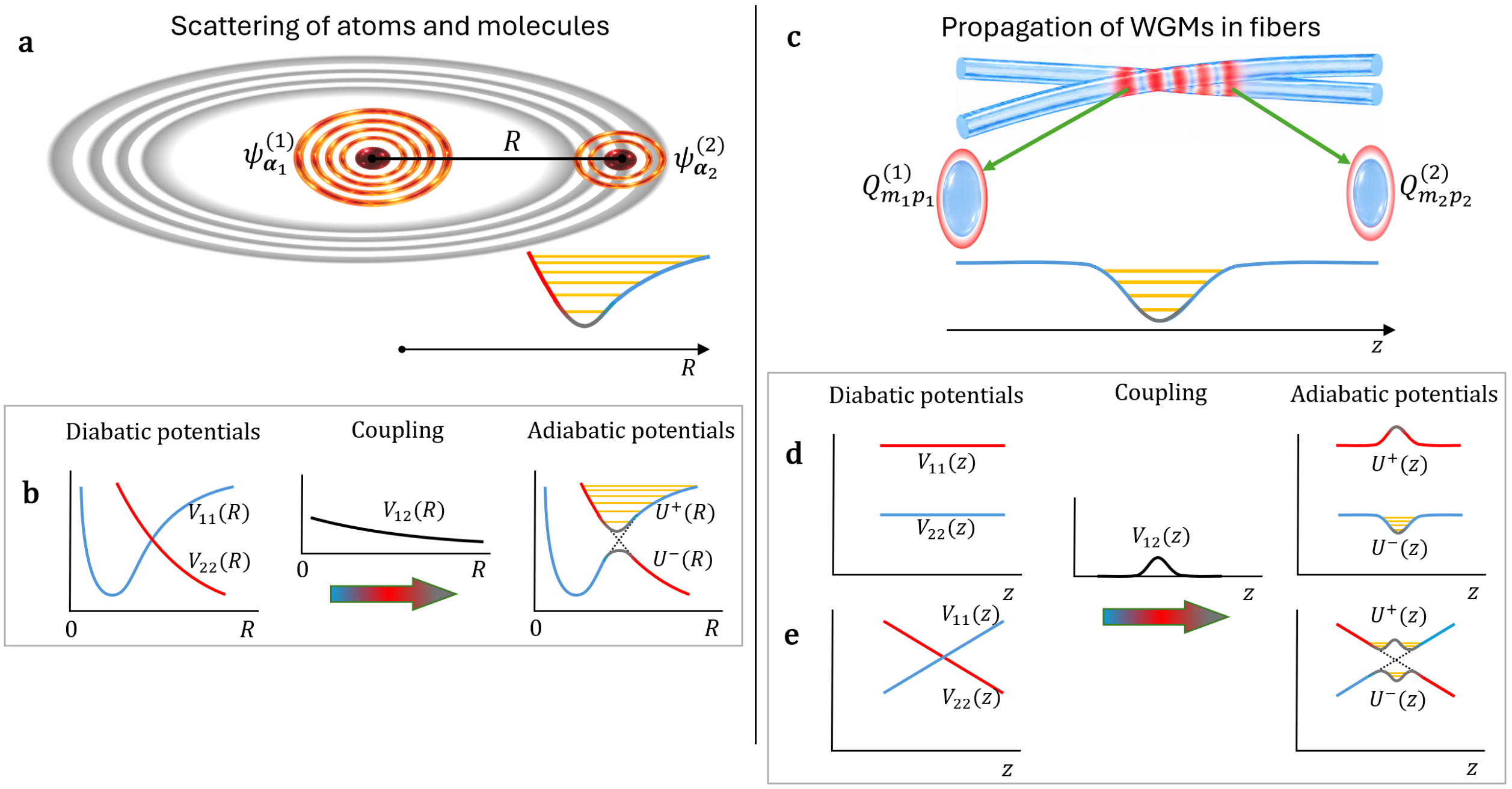

**Fig. 1.** System configurations and characteristic profiles of diabatic and adiabatic potentials for the coupled-state atomic or molecular scattering and WGM propagation in side-coupled fibres. **a** Illustration of atomic or molecular scattering. **b** Crossing diabatic, coupling, and adiabatic potentials. **c** Illustration of side-coupled fibres and localized WGM propagation. **d** Coupling-induced diabatic and adiabatic potentials for noncrossing cutoff frequency profiles. **e** Coupling-induced diabatic and adiabatic potentials for intersecting cutoff-frequency profiles. For weaker coupling or steeper slopes of potentials the avoided crossing configuration shown in **e** is similar to that in **b**.

Here we note that the slow axial propagation of whispering-gallery modes (WGMs) in side-coupled optical fibres provides a particularly versatile setting for exploring photonic analogues of quantum-mechanical avoided crossing. In an uncoupled optical fibre, these WGMs circulate near the fibre surface while propagating slowly along its axis. The WGM axial propagation is governed by a one-dimensional

Schrödinger-type equation, in which the local cutoff frequency – the frequency at which the axial propagation constant vanishes – plays the role of an effective potential [18]. When two fibres are side coupled, their WGMs hybridize, and the spatially varying inter-fibre coupling can itself create a microresonator localized near the coupling region [19–22] (Fig. 1c). Unlike the coupling $V_{12}(R)$ commonly treated as constant or slowly varying near an isolated atomic or molecular avoided crossing (Fig. 1b), the inter-fibre WGM coupling $V_{12}(z)$ is strongly localized along the fibre axis $z$ (Figs. 1d and 1e). It can therefore reshape the *adiabatic potentials*, $U^{+}(z)$ and $U^{-}(z)$, much more strongly and produce richer spatial and spectral behaviour. In addition to the coupling-modified profiles shown in Fig. 1b, these adiabatic potentials can acquire coupling-dominated well and barrier profiles such as those illustrated in Fig. 1d, as well as the double-well and double-barrier profiles shown in Fig. 1e.

In previous realizations of side-coupling-induced fibre microresonators, reconfigurability was achieved by bending [19, 22], mutual rotation [20, 22], or relative translation [21] of optical fibres. These operations primarily modified the axial profile of the inter-fibre coupling, thereby tuning the microresonator dimensions and free spectral range. However, these systems could not be tuned controllably toward the cutoff-frequency avoided crossing, where coupling-induced hybridization is strongest. Operation near this configuration in Refs. [19–21] therefore relied on selecting, from a long parent fibre, segments with suitably matched cutoff frequencies.

Mechanical stretching provides precisely this missing degree of freedom. Here, we use it to tune the relative cutoff frequencies of the two fibres through the avoided-crossing region and thereby control both the shape and spectrum of the side-coupling-induced microresonator. Stretching has long been used for precise WGM tuning in monolithic sphere, bottle, bubble, and capillary microresonators [23–30]. *In those devices, however, strain acts directly on the entire resonator body, shifting the spectrum of a cavity whose confinement mechanism is already established.*

In the coupled fibres configuration such as illustrated in Fig. 2a *stretching plays a qualitatively different role*. Applied only to the straight fibre, it shifts the uncoupled cutoff frequency of that fibre relative to that of the bent fibre while leaving the localized coupling geometry essentially unchanged. It thereby provides direct control over the cutoff-frequency detuning shown in Figs. 2b without altering the axial coupling profile. *Sweeping this detuning through the avoided crossing reshapes the effective axial potential and the localized states it supports*. As described in the next section, this evolution becomes particularly striking when the microresonator is probed with a microfibre coupled to the straight fibre (Fig. 2a). Its spectrum detaches from the cutoff-frequency trace of that fibre and appears to "hang" below it (Fig. 2c). *Stretching, therefore, controls the formation, spectral position, height, and confinement strength of the coupling-induced composite microresonator, rather than merely shifting the resonances of an already established monolithic cavity*.

The paper is organized as follows. In the section "Avoided crossing of WGM cutoff frequencies," we discuss the physical effects accompanying the tuning of the straight fibre's cutoff frequency through that of the bent fibre, as illustrated in Fig. 2. In the section "Effect of surrounding WGMs", we investigate how additional fibre modes influence the behaviour of two WGMs with closely spaced cutoff frequencies. In the section "Effect of fibre stretching", we discuss how stretching the fibre shifts WGM cutoff frequencies. In the section "Experimental results", we describe the experimental setup and present the observed spectral evolution under stretching through the cutoff-frequency avoided-crossing region. In section "Adiabatic perturbed two-mode fitting", we show that the experimental observations are in remarkably good agreement with the developed theoretical model. Finally, in "Conclusions", we summarize the main findings of this work.

# Results

## Avoided crossing of WGM cutoff frequencies

In general, an optical fibre supports multiple WGMs, each characterized by its own cutoff frequency. The interaction of side-coupled fibres is therefore intrinsically a multichannel problem, even when a particular avoided crossing is dominated by two nearly resonant modes. The general $N$-channel formulation is thus the appropriate starting point. For the configurations considered below, the general equations can be simplified by applying the adiabatic approximation, retaining the interaction between the two nearly resonant WGMs exactly, and treating the influence of the remaining modes perturbatively.

Let $\Psi_n(z)$ denote the slowly varying axial amplitude of channel $n$, whose uncoupled local cutoff frequency is $\nu_{c,n}(z)$. We also introduce optical frequency $\nu$, a reference frequency $\nu_0$, matrix potential $\mathbf{V}(z)$ and energy in frequency units $E = \nu - \nu_0$. The diagonal elements of matrix potential $\mathbf{V}(z)$ are *diabatic potentials*

$$V_{nn}(z) = \nu_{c,n}(z) - \nu_0, \tag{1}$$

while its nondiagonal elements are the coupling coefficients $V_{mn}(z)$ between fibre $m$ and $n$. Following the standard coupled-mode and local-mode description of weakly interacting optical waveguides [31] and its application to side-coupled WGM fibres [22], we arrive at the coupling mode equations for the axial amplitudes $\Psi_n(z)$:

$$\begin{gathered}\left(-\eta\mathbf{I}\frac{d^2}{dz^2} + \mathbf{V}(z)\right)\Psi(z) = E\Psi(z),\\ \Psi(z) = \left(\Psi_1(z),\ldots,\Psi_N(z)\right)^{\mathsf{T}},\ \mathbf{V}(z) = \{V_{mn}(z)\},\ \eta = \frac{c^2}{8\pi^2 n_0^2 \nu_0}.\end{gathered} \tag{2}$$

Here $\mathbf{I}$ is the $N \times N$ unity matrix, $c$ is the speed of light, and $n_0$ is the fibre refractive index. To follow the analogy to quantum mechanics [14, 15] and following Eq. (1), we will call the cutoff frequencies $\nu_{c,n}(z)$ the *diabatic potentials* governing the WGM propagation.

In the case of a single fibre, Eq. (2) is simplified to the one-dimensional Schrödinger-type equation commonly used in the SNAP theory [22],

$$\left(-\eta\frac{d^2}{dz^2} + V_{11}(z)\right)\Psi(z) = E\,\Psi(z) \tag{3}$$

Importantly, unlike SNAP microresonators [18], which are formed by nanoscale axial variations of the effective fibre radius and hence of the cutoff frequency and potential $V_{nn}(z)$, the formation of fibre-coupling-induced localized states and microresonators do not require the individual-fibre potentials to vary with $z$. The coupled fibres may be uniform, with $V_{nn}(z) = \text{const}$, while localization can arise solely from the axial variation of the inter-fibre coupling $V_{mn}(z)$.

Of major interest for us in this paper will be the solution of Eq. (2) in the *adiabatic approximation* when the going forward and backward solutions are searched as $\Psi_n(z) = A_n(z)\exp\left(\pm i\int^z \beta(z',E)\,dz'\right)$ with slow varying preexponent $A_n(z)$ and propagation constant $\beta(z,E)$. Neglecting derivatives of $A_n(z)$ and $\beta(z,E)$ at leading adiabatic order reduces Eq. (2) to the system of linear algebraic equations

$$\sum_{m=1}^{N}\left[V_{nm}(z) + \left(\eta\beta^2(z,E) - E\right)\delta_{nm}\right]A_m(z) = 0, \quad n = 1,\ldots,N. \tag{4}$$

In the photonic counterpart shown in Fig. 1c, we assume $N = 2$ and the two diabatic channels are WGMs of the separate fibres, characterized schematically by the transverse modes $Q_{m_1p_1}^{(1)}$ and $Q_{m_2p_2}^{(2)}$. The uncoupled cutoff-frequency profiles $V_{11}(z)$ and $V_{22}(z)$ are the diabatic potentials. Diagonalization of the potential matrix in Eq. (4) gives the *adiabatic potentials*

$$U^{\pm}(z) = \frac{V_{11}(z) + V_{22}(z)}{2} \pm \sqrt{\left(\frac{V_{11}(z) - V_{22}(z)}{2}\right)^2 + |V_{12}(z)|^2}. \tag{5}$$

This equation is mathematically analogous to that used in the theory of atomic and molecular collisions [14, 15]. However, comparison of Figs. 1b, 1d, and 1e reveals an important physical difference. Near an isolated avoided crossing in atomic or molecular scattering, the coupling $V_{12}(R)$ is commonly approximated as constant or slowly varying over the relevant interval of atomic or molecular separation $R$ [14, 15]. By contrast, in the coupled-fibre system, $V_{12}(z)$ is strongly localized near the fibre-contact region and directly reshapes the adiabatic potentials.

To better understand the effect of this localized coupling, it is instructive to apply Eq. (5) to the case of two axially uniform fibres with diabatic potentials (cutoff frequencies) $V_{11}(z) = V_{11}$ (fibre 1) and $V_{22}(z) = V_{22}$ (fibre 2) which are independent of $z$. We assume that fibre 1 is straight and subjected to axial stretching.

Fibre 2 is bent, remains coplanar with fibre 1, and touches it at the axial position $z = 0$ (Fig. 2a). We model the coupling coefficient $V_{12}(z)$ by a Gaussian function of the axial coordinate $z$:

$$V_{12}(z) = W_0 \exp\left(-\frac{z^2}{z_0^2}\right). \tag{6}$$

In the semiclassical approximation for weak inter-fibre WGM coupling, parameters $z_0$ and $W_0$ of this equation were found in Ref. [22] as

$$z_0 = \frac{c^{1/2}R^{1/2}}{\pi^{1/2}(n_0^2-1)^{1/4}\nu_0^{1/2}}, \quad W_0 = \frac{c^{3/2}}{2^{3/2}\pi^2(n_0^2-1)^{3/4}\nu_0^{1/2}r_0^{3/2}}. \tag{7}$$

Here $r_0$ is the radii of fibres assumed equal and $R$ is the curvature radius of fibre 2. For the fibres used in our experiments having $r_0 = 40$ μm, a comparison of this approximation with numerical results obtained using the multipole method [22] is presented in Fig. 5 below. In Figs. 2b, diabatic potential $V_{11}$ of fibre 1 increases from left to right, while the diabatic potential $V_{22}$ of the stationary fibre 2 remains fixed. From Eqs. (5) the localized coupling produces a potential well in $U^-(z)$ and a corresponding barrier in $U^+(z)$. These potentials are presented in Figs. 2b and 2c in the dimensionless form in units of $W_0$ as a function of axial distance in units of $z_0$.

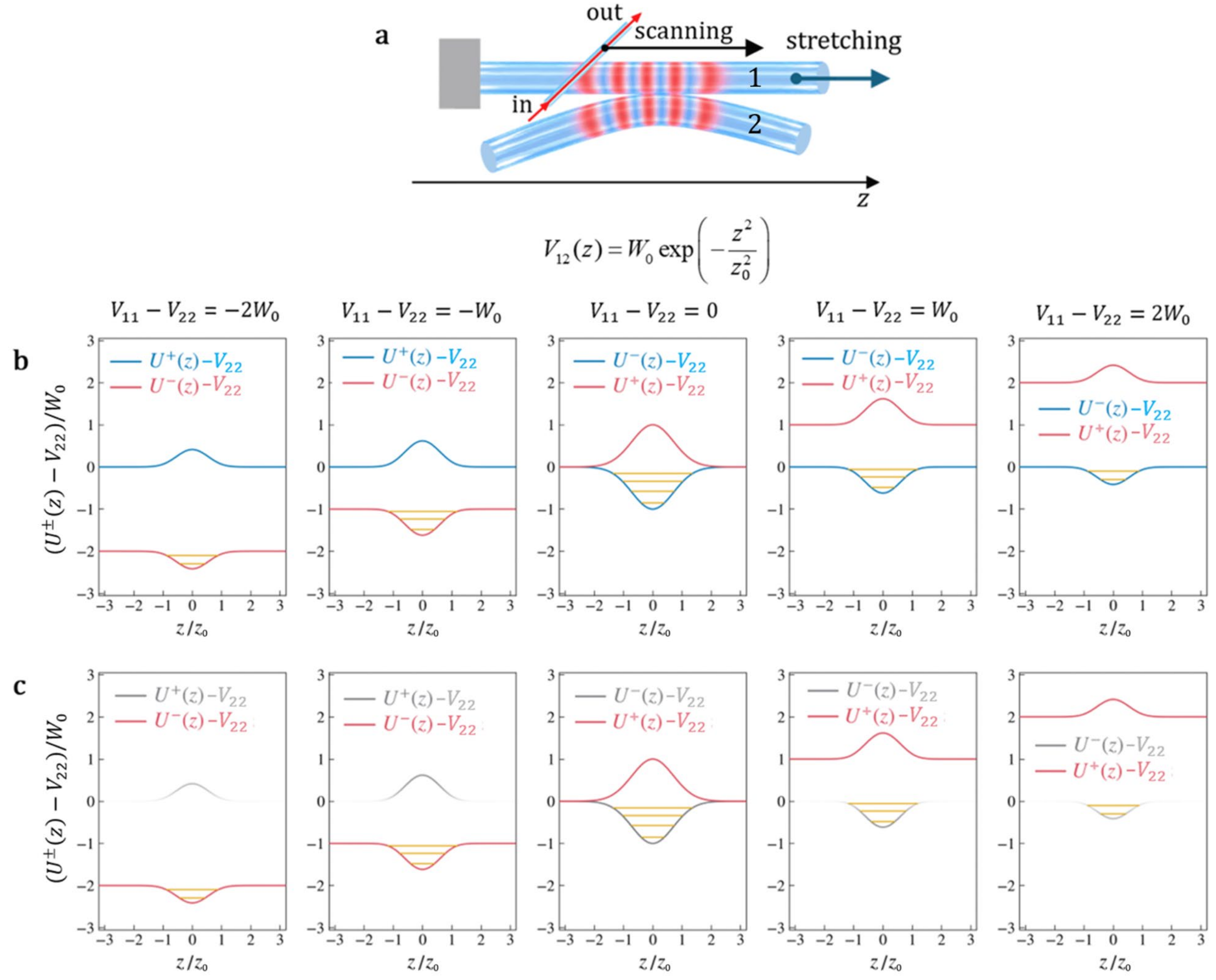

**Fig. 2.** Evolution of the coupled potentials during a cutoff-frequency sweep. **a** Straight stretching fibre coupled to a bent fibre and to an input-output microfibre. **b** and **c** From left to right: the diabatic potential detuning passes from one side of the crossing through zero to the other side. Horizontal orange lines indicate discrete eigenfrequencies supported by the coupling-induced potential well. In the lower row, the complementary coupled potential is de-emphasized in gray to highlight the localized well and its eigenfrequencies visible from the side of the stretching fibre and input-output microfibre.

The depth of the potential well $U^{-}(z)$ and the height of the barrier in $U^{+}(z)$ in Fig. 2b have the same value, which achieves a maximum at the avoided crossing $V_{11} - V_{22} = 0$. As the crossing is approached, the well deepens and accommodates an increasing number of localized axial states, shown by the horizontal yellow lines in Fig. 2b. Away from the crossing, the well becomes shallower and its spectrum contracts. Consequently, for $V_{11} < V_{22}$, the localized eigenfrequencies move upward as the crossing is approached. However, for $V_{11} > V_{22}$, the continuum threshold of $U^{-}(z)$ becomes fixed by $V_{22}$. The localized eigenfrequencies then remain near the cutoff frequency of the bent fibre even though the cutoff frequency of the straight fibre continues to increase.

Figures 2c represents the same evolution as in Figs. 2b observed through the microfibre coupled to the straight fibre 1. Before the crossing, the localized states of $U^{-}(z)$ have the stretched-fibre character and therefore follow its diabatic potential trace. Beyond the crossing, $U^{-}(z)$ becomes associated with the bent fibre 2, although its fibre 1 component keeps its localized states observable. Their frequencies no longer follow the rising cutoff frequency $V_{11}$ of the straight fibre 1 but remain close to the lower, nearly stationary continuum threshold at $V_{22}$. This change of modal character produces the apparent "hanging" of the microresonator spectrum in Fig. 2c. With increasing detuning beyond the crossing, the depth of the potential well decreases, accounting for the contraction of the observed spectrum.

## Effect of surrounding WGMs

The two-mode description of the avoided crossing can be extended to account for the weak influence of the surrounding WGMs. We retain the two selected modes, labelled 1 and 2, as an explicitly coupled pair and treat their coupling to the remaining modes, labelled $n = 3, \dots, N$, perturbatively. This reduction is valid when the coupling to each remote mode is weak compared with its cutoff-frequency separation from the selected modes throughout the relevant axial interval. Eliminating the remote-mode amplitudes to second order gives an effective Hermitian two-mode matrix, which is then diagonalised without expanding the direct coupling within the selected pair. Then, as shown in Supplementary Information, Section S1, Eq.(5) retains its form provided that the matrix elements $V_{ij}(z)$ are replaced by their effective values $V_{ij}^{\text{eff}}(z)$:

$$V_{jj}^{(eff)}(z) = V_{jj}(z) + \sum_{n=3}^{N} \frac{|V_{jn}(z)|^2}{V_{jj}(z) - V_{nn}(z)}, \quad j = 1,2,$$
$$V_{12}^{(eff)}(z) = V_{12}(z), \quad V_{21}^{(eff)}(z) = V_{21}(z). \tag{8}$$

All quantities on the right-hand side of these equations are elements of the original $N$-channel matrix in Eq. (4). Each diagonal correction is a sum of shifts caused by coupling to remote WGMs: modes with higher cutoff frequencies give negative contributions, whereas modes with lower cutoff frequencies give positive contributions. These corrections can therefore reshape the adiabatic potential profiles. In the model considered here, coupling connects WGMs on different fibres, while the uncoupled modes on each fibre form orthogonal channels. Since the two selected modes belong to different fibres, each remote mode can couple to at most one of them, so $V_{1n}(z)V_{n2}(z) = 0$. There is consequently no second-order correction to the off-diagonal coupling between the selected modes, i.e., $V_{j,3-j}^{(eff)}(z) = V_{j,3-j}(z)$, where $j = 1,2$.

Determining the actual axial profiles from the nominal fibre geometry alone is difficult. The individual cutoff frequencies depend on small axial variations of the fibre radius and refractive index, as well as on strain and bending. The inter-fibre coupling is sensitive to the local separation, mutual alignment and curvature near the contact region, where mechanical deformation can modify the nominal geometry. Calculating the second-order corrections additionally requires the cutoff frequencies and field overlaps of all appreciably contributing remote WGMs. Therefore, below we determine the parameters describing the effective axial profiles by comparison with the experimentally measured cutoff-frequency traces and induced resonator spectra.

## Effect of fibre stretching

Axial stretching tunes the diabatic cutoff frequency of the straight fibre through the combined effects of radial contraction and the photoelastic response of silica. For a WGM of fixed azimuthal and radial order, both effects reduce the optical path length around the fibre circumference and therefore shift the cutoff frequency upwards. In the limit of small, uniform strain, the shift is related to the fibre elongation by (see Supplementary Information, Section S2) [32, 33]

$$\frac{\Delta \nu_{c,q}}{\nu_{c,q}} = C_q \frac{\Delta L}{L}, \quad C_q = \mu_{\mathrm{P}} + p_{e,q}. \tag{9}$$

Here, $L$ is the length of the uniformly strained silica segment, $\Delta L$ is its actual elongation, $\mu_{\mathrm{P}}$ is the Poisson ratio and $p_{e,q}$ is the effective photoelastic coefficient for polarisation $q$. For a silica fibre, we have $C_{\mathrm{TE}} = 0.369$ and $C_{\mathrm{TM}} = 0.201$ (see Supplementary Information, Section S2). For example, a TE-like WGM near 193 THz, a 100 GHz sweep requires relative elongation of $\Delta L/L \cong 1.4 \times 10^{-3}$. For the 55 mm segment used in our experiments, this shift corresponds to an elongation of approximately 77 μm. Consequently, a tuning resolution of 130 MHz can be achieved for precision translation stages with a typical 0.1 μm displacement resolution.

For a bare silica fibre under uniform axial stress, the required force is (Supplementary Information, Section S2),

$$F = \pi r_0^2 Y \frac{\Delta L}{L} = \frac{\pi r_0^2 Y}{C_q} \frac{\Delta \nu_{c,q}}{\nu_{c,q}}, \tag{10}$$

where $Y$ is Young's modulus (72 GPa for silica) and $r_0$ is the fibre radius. For a silica fibre of radius $r_0 = 40$ μm used in our experiment described below, 100 GHz shift requires approximately 0.51 N pulling force.

## Experimental results

In our experiment, two silica fibre segments with radius $r_0 = 40$ μm and refractive index $n_0 = 1.44$ (Thorlabs, SM1500G80) were used. After removing the coating, the fibre segments were cleaned with methanol and flame treated. Fibre 1 was kept straight, while fibre 2 was bent and arranged as shown in Fig. 3. The straight fibre was mounted on a motorized translation stage, enabling controlled axial stretching. The fibres were gently brought into contact to form a coupling region. Light was launched into fibre 1 (the straight fibre) through a transverse input–output microfibre taper connected to a Luna OVA-5000 Optical Vector Analyzer. As illustrated in Fig. 2a, the microfibre was translated along the axis of the straight fibre 1 and brought into contact with it at 5 μm intervals. A transmission spectrum was recorded at each contact position. The measured data were displayed as spectrograms of the transmission power, $P(z, \nu)$, as a function of the optical frequency $\nu$ and the axial position $z$ of the microfibre.

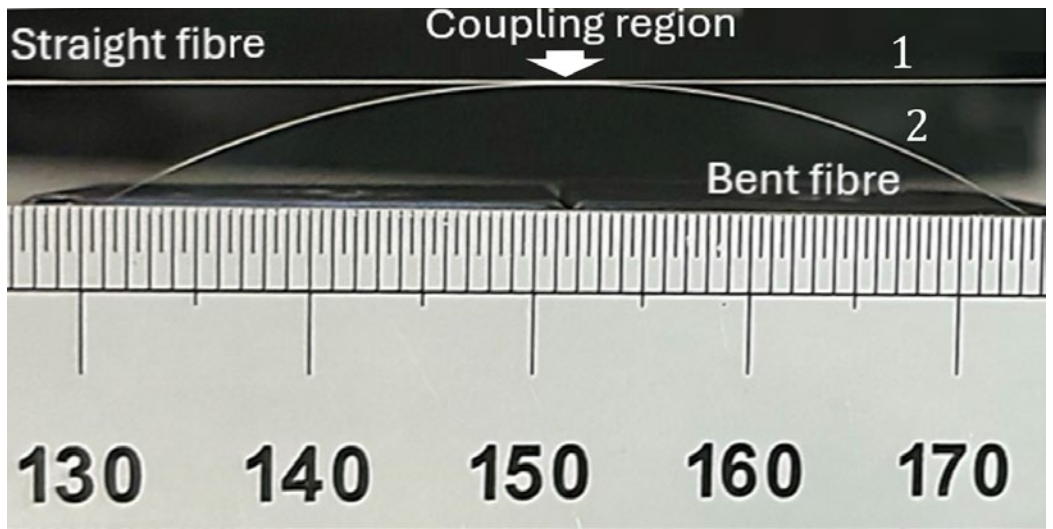

**Fig. 3.** Experimental arrangement of straight fibre 1 and bent fibre 2.

To investigate the stretching-induced tunability of the microresonators formed by coupling between the straight and bent fibres, the straight fibre was elongated by 20 μm using a motorised translation stage. The resulting elongation shifted its cutoff frequencies (diabatic potentials), as described above. In this experiment, the straight and bent fibre segments were selected from the same long parent fibre so that the chosen cutoff frequency of the straight fibre lay slightly below a cutoff frequency of the bent fibre. Stretching the straight fibre then tuned its diabatic potential through that of the bent fibre, revealing the avoided crossing of the coupled-mode cutoff frequencies described above.

Figure 4a shows the spectrogram near the selected cutoff frequency of bent fibre 2 in the absence of coupling to fibre 1. Figure 4b shows the spectrogram of the microresonator formed by coupling the two fibres, with fibre 1 still unstretched [19]. Complementing the original cutoff-frequency profile of fibre 2 shown in Fig. 4a, the corresponding profile of uncoupled fibre 1 can be revealed from Fig. 4b by a straight-line interpolation between its values in the uncoupled regions on either side of the coupling region. From Figs. 4a and 4b, we find the nonuniformity of fibres 1 and 2 as 27 GHz/mm and 17 GHz/mm, which corresponds to their effective radius variations 5.6 nm/mm and 3.5 nm/mm, respectively. These nonuniformities are much greater than the characteristic nonuniformities ~ 0.1 nm/mm of commercial single mode optical fibres with radius $r_0 = 62.5$ μm [34].

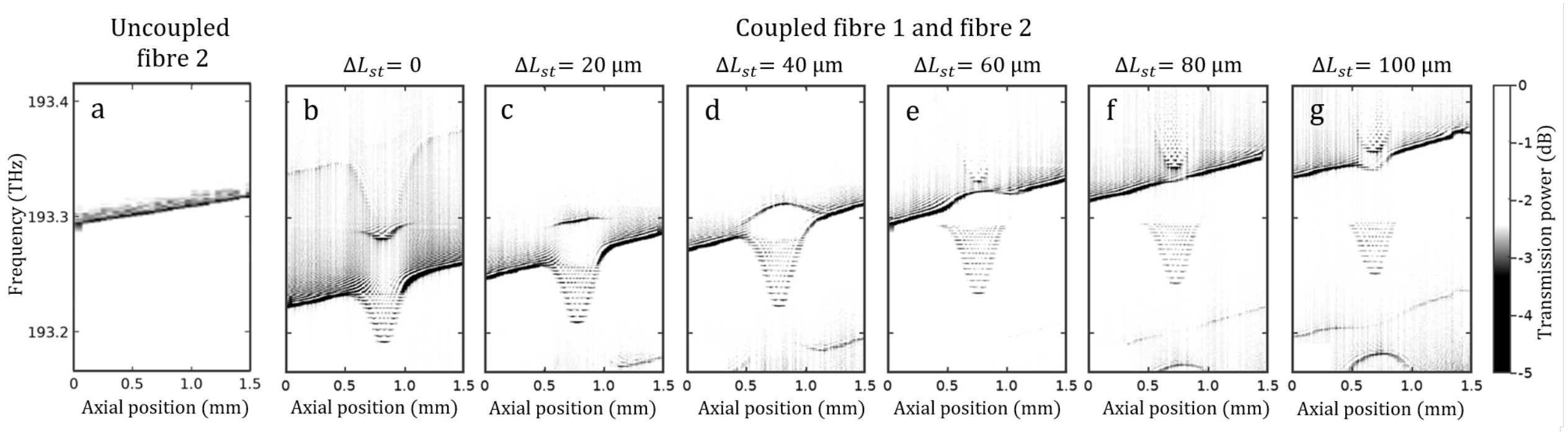

**Fig. 4. a** A spectrogram of uncoupled bent fibre 2 near the cutoff frequency considered. **b**-**g** Spectrograms of microresonators $P(z,\nu)$ induced along the fibre coupling region where the straight fibre was gradually stretched with a pulling length increment of 20 μm for each case.

Figures 4c–g show the spectrograms recorded as the stretch-induced elongation of fibre 1 (original length $L$ = 55 mm) was increased from $\Delta L_{st}$ = 20 μm to 100 μm in 20 μm increments. The colour scale was chosen to enhance the visibility of the resonant features. At the left- and right-hand sides of the spectrograms, coupling between the fibres is negligible. Within the coupling region, the local distortion of the cutoff-frequency profiles indicates the formation of an optical potential well supporting microresonator modes. The observed evolution is qualitatively similar to the theoretical results in Fig. 2: the well initially deepens as the selected cutoff frequencies approach one another, reaches its greatest depth near minimum detuning, and becomes shallower as further stretching increases their detuning. At larger elongations, the discrete resonances lie below and become spectrally separated from the cutoff-frequency profile of fibre 1, corresponding to the hanging microresonators illustrated in Fig. 2c. These results demonstrate continuous mechanical tuning of the microresonator depth. However, unlike the constant diabatic potentials assumed in Fig. 2, the experimental cutoff-frequency profiles are tilted along the fibre axis. Quantitative fitting therefore requires a separate analysis accounting for these axial slopes and perturbations from surrounding WGMs presented in the following section.

## Adiabatic perturbed two-mode fitting

We now apply the perturbed adiabatic two-mode approximation to the experimental spectrograms in Fig. 4. The influence of surrounding modes depends on their spectral detuning and field overlap with the selected WGMs. Here we assume that modes with considerable overlap can be taken into account through the second-order corrections described by Eq. (8). We first fit the observed cutoff-frequency dependencies on the microfibre position (Fig. 4) with this model and then assess whether the fitted behaviour is consistent with our physical assumptions.

In the perturbed two-mode approximation described above, for each fibre 1 elongation $\Delta L_{st}$, the fitted adiabatic potentials are expressed through the effective diabatic potentials and inter-mode coupling as

$$U^{\pm}(z,\Delta L_{st}) = \frac{V_{11}^{(eff)}(z,\Delta L_{st}) + V_{22}^{(eff)}(z,\Delta L_{st})}{2} \pm \sqrt{\frac{\left(V_{11}^{(eff)}(z,\Delta L_{st}) - V_{22}^{(eff)}(z,\Delta L_{st})\right)^2}{4} + \left(V_{12}(z,\Delta L_{st})\right)^2} \quad . \quad (11)$$

To include the measured axial nonuniformity of fibres and the local perturbations from surrounding WGMs, we set

$$V_{11}^{(eff)}(z,\Delta L_{st}) = a_1(\Delta L_{st}) + b_1 z + A_1(\Delta L_{st})G^2(z - z_c(\Delta L_{st})) - \nu_0$$
$$V_{22}^{(eff)}(z,\Delta L_{st}) = a_2(\Delta L_{st}) + b_2 z + A_2(\Delta L_{st})G^2(z - z_c(\Delta L_{st})) - \nu_0 \quad (12)$$
$$V_{12}(z,\Delta L_{st}) = W_0 G(z - z_c(\Delta L_{st}))$$

Here $a_j(\Delta L_{st})$ are the uncoupled cutoff frequency of fibres with $j = 1,2$ at $z = 0$ in the axial coordinate system used in Fig. 4 and $b_j$ are their axial slopes. The measurements of uncoupled fibres give approximately $b_1$=27 GHz/mm and $b_2$=17 GHz/mm. The dependence $z_c(\Delta L_{st})$ allows for a small displacement of the coupling region for different $\Delta L_{st}$. The coefficients $A_j(\Delta L_{st})$ in Eq. (12) can be expressed through the amplitudes of the cutoff-frequency shifts induced by coupling to surrounding WGMs. Indeed, as shown in Ref. [22], in semiclassical approximation, the coupling coefficients $V_{jn}(z)$ in Eq. (8) are proportional to one another, i.e., $|V_{jn}(z)|^2 \simeq |V_{jn}(z_c)|^2 G^2(z\text{-}z_c)$. Then, after introducing the dependence on stretching $\Delta L_{st}$, Eq. (8) gives

$$V_{jj}^{(\mathrm{eff})}(z,\Delta L_{st}) = V_{jj}(z,\Delta L_{st}) + A_j(\Delta L_{\mathrm{st}})\, G^2(z\text{-}z_c(\Delta L_{st})),$$
$$A_j(\Delta L_{st}) = \sum_{n=3}^{N} \frac{|V_{jn}(z_c(\Delta L_{st}))|^2}{V_{jj}(z_c(\Delta L_{st})) - V_{nn}(z_c(\Delta L_{st}))}, \quad j=1,2. \quad (13)$$

Consequently, the direct coupling varies as $G$, while its second-order diagonal correction varies as $G^2$, justifying the expressions in Eq. (12). It follows from Eq. (13) that the signs of $A_j(\Delta L_{st})$ are set by the spectral positions and overlap strengths of the remote modes. To fit the spectrograms in Fig. 4, we assume a Gaussian form for $G(z)$ in Eq. (12):

$$G(z) = \exp\left(-\frac{z^2}{z_0^2}\right). \quad (14)$$

Consequently, $W_0$ in Eq. (12) and $z_0$ in Eq. (14) determine the amplitude and axial width of the direct coupling between the two selected WGMs. This approximation is supported by the comparison in Fig. 5 of the axial coupling profiles obtained analytically (Eqs. (6) and (7)), numerically, and by fitting experimental spectrograms. In Fig. 5, analytical and numerical profiles are found for the fibres with radius $r_0$=40 $\mu$m considered in our experiments. In our numerical calculations using the multipole method [22], we consider the fundamental radial WGM with azimuthal and radial quantum numbers $(m,q)$=(223,1), near frequency 193 THz. The analytical and numerical couplings are found with a relatively good agreement, with maxima equal to 50.2 and 55.1 GHz and the full widths at half maximum equal to 0.257 and 0.251 mm, respectively.

Figure 6a compares the reconstructed uncoupled cutoff-frequency profiles $a_2(\Delta L_{st}) + b_2 z$ of bent fibre 2 found from Eq. (12) with its independently measured spectrogram (Fig. 4a). The profiles are obtained by fitting the six coupled-fibre spectrograms in Figs. 4b–g. Their small spread reflects the weak dependence of the fitted cutoff frequency of fibre 2 on the elongation of fibre 1.

The remarkable agreement between the fitted cutoff-frequency profiles and the experimental spectrograms shown in Figs. 6b–g further supports the Gaussian approximation in Eq. (14). The fitted Gaussian profile has a peak of approximately 41.6 GHz and a full width at half maximum of about 0.32 mm. From Eq. (7), its width corresponds to an effective radius $R_{fit}$=78.6 mm. This value is significantly larger than $R \simeq 45 - 50$ mm estimated from the photograph in Fig. 3. We suggest that this difference may be caused by slight elastic deflection of the nominally straight fibre or deformation of the bent fibre near contact that can broaden the region of close proximity yielding a larger effective radius. Such deformation may be promoted by van der Waals adhesion, electrostatic attraction [35, 36] or capillary forces from adsorbed moisture [37] as well as additional pushing after contact. Figure 5 also shows that the maximum of the fitted coupling is 25% smaller than its numerical prediction. The latter can be caused by surface roughness, residual contamination, and by limitations of the adopted perturbed two-mode approximation.

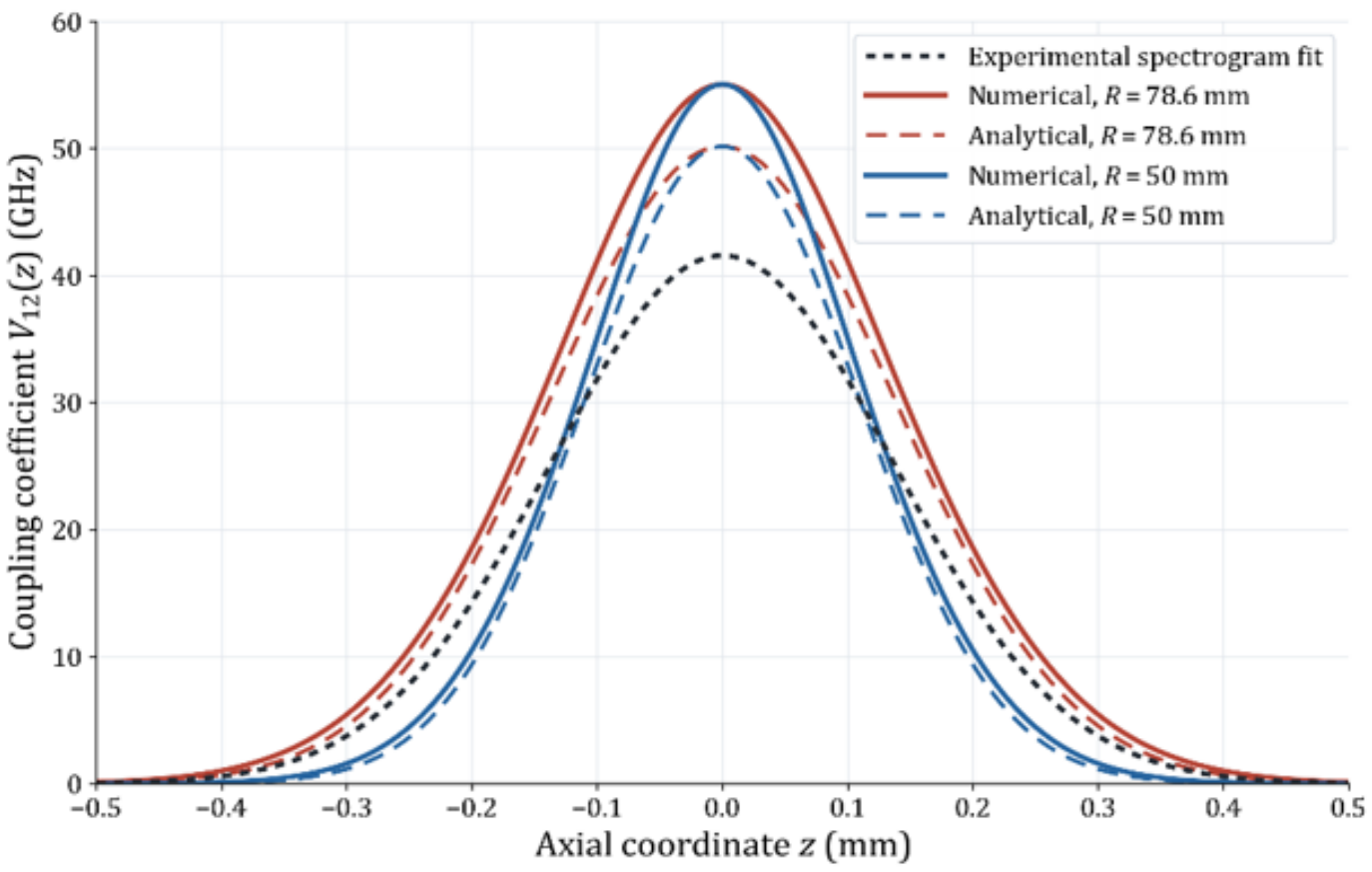

**Fig. 5.** Comparison of the Gaussian coupling obtained from the experimental spectrogram fit (black dotted curve) with numerical multipole calculations (solid curves) and analytical calculations (dashed curves). Red and blue curves correspond to effective bend radii of 78.6 mm and 50 mm, respectively. All curves use the off-diagonal coupling convention, equal to half the real frequency splitting of the tracked resonant doublet.

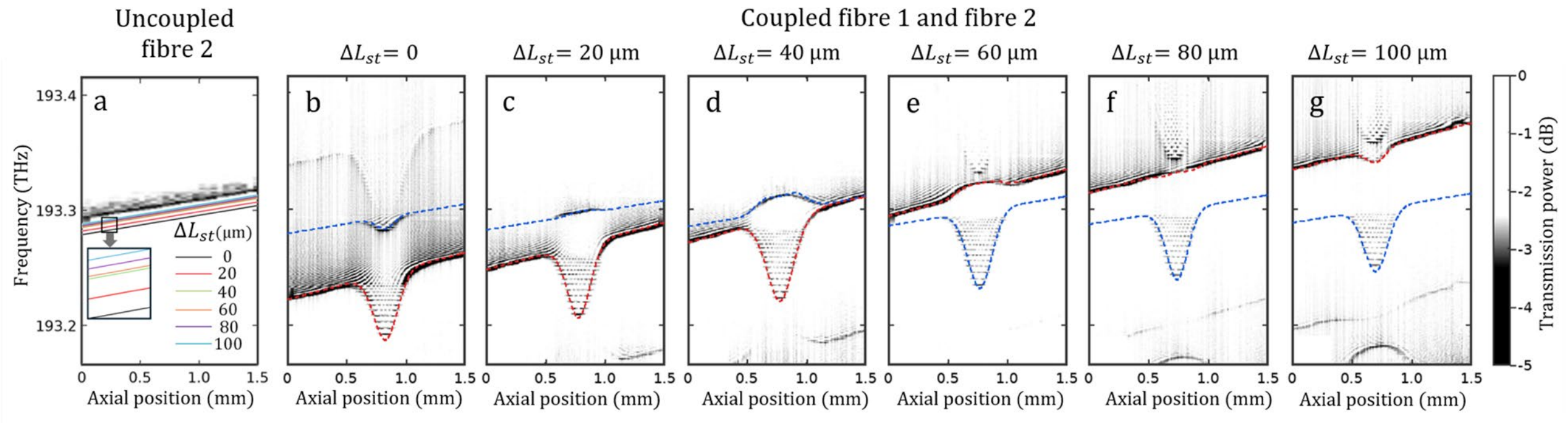

**Fig. 6.** Fitted cutoff-frequency profiles compared with the measured spectrograms. **a** Reconstructed uncoupled backgrounds of bent fibre 2 superimposed on its isolated-fibre spectrogram. **b–g** Calculated adiabatic cutoff-frequency trajectories superimposed on the spectrograms for successive elongations of fibre 1. Red and blue distinguish branches associated, away from the coupling region, with straight fibre 1 and bent fibre 2, respectively.

The fitted uncoupled cutoff frequencies $a_j(\Delta L_{st})$ at $z = 0$, shown in Fig. 7a, increase nearly linearly with elongation for the stretched fibre 1 and exhibit only a small drift for the bent fibre 2. The fitted tuning rate is in good agreement with the prediction of approximately 1.30 GHz/μm obtained from Eq. (9) for a TE-like WGM near 193 THz and fibre length $L$=55 mm. Minor departures from the linear dependence of parameters $a_1(\Delta L_{st})$ for fibre 1 and a small drift of parameters $a_2(\Delta L_{st})$ for fibre 2 are presumably due to the compliance or small motions in the mounts.

The fitted amplitudes $A_j(\Delta L_{st})$ in Fig. 7b vary approximately antisymmetrically about a common negative offset. Equation (13) provides a possible explanation for this behaviour. Writing the remote-mode detunings as $D_{jn} = V_{jj} - V_{nn}$, small changes at approximately fixed overlap amplitudes give $\delta A_j \simeq -\sum_n \left|V_{jn}\right|^2 \delta D_{jn}/D_{jn}^2$. Stretching raises the cutoff frequencies on fibre 1, increasing the detunings for its coupling to modes on fibre 2 and decreasing those for the reciprocal contributions to fibre 2. Comparable overlap-weighted contributions can therefore produce approximately opposite changes in the two correction amplitudes. Although these contributions vary monotonically within an interval containing no remote-mode crossing, crossings between sampled elongations can produce a nonmonotonic sequence of fitted amplitudes. Such intervening crossings do not exclude a perturbative description of the measured spectrograms, provided that the contributing remote modes remain sufficiently detuned from both retained branches throughout the coupling region at each measured setting. Near an intervening crossing, however, the additional mode would need to be included explicitly.

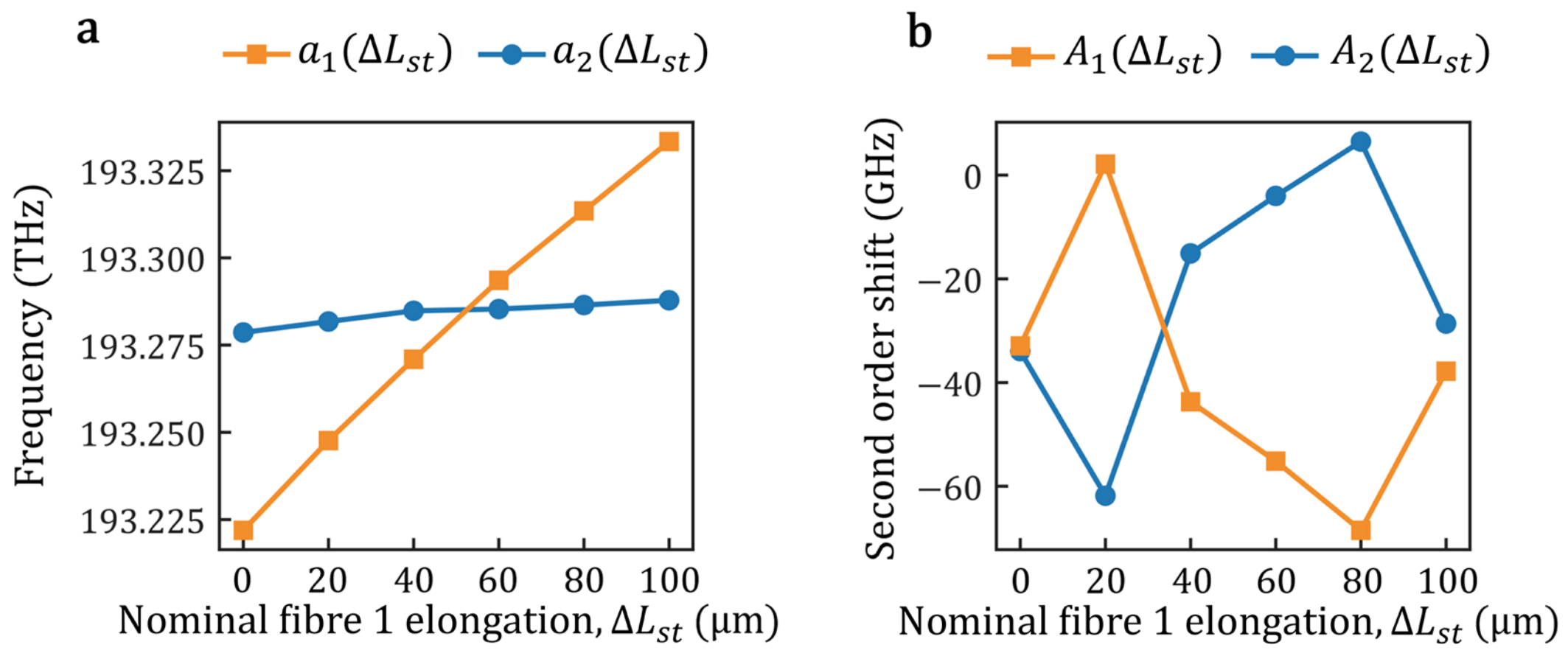

**Fig. 7.** **a** The fitted cutoff-frequency values $a_j(\Delta L_{st})$ at a fixed input-output microfibre position and **b** the local second-order correction amplitudes $A_j(\Delta L_{st})$ as functions of the elongation $\Delta L_{st}$ of fibre 1.

Overall, the agreement between the fitted and measured coupled cutoff-frequency profiles, the nearly antisymmetric behaviour of the diagonal corrections and the expected stretching trend suggest that the perturbed two-mode model provides a satisfactory description of the experiments. More precisely, its validity requires coupling to remote modes to be weak relative to their detunings from both retained branches, with a small combined contribution of these modes to the coupled eigenmodes. The adiabatic reduction additionally requires sufficiently slow axial variation to suppress conversion between the local branches. If one or more additional WGMs approach either retained branch closely enough for their detunings to become comparable to their coupling strengths, those modes must be included explicitly in the model.

## Conclusions

In conclusion, we have demonstrated stress-controlled evolution of a WGM microresonator formed by side coupling between straight and bent optical fibres. By mechanically stretching the straight fibre while maintaining the coupling geometry approximately unchanged, we continuously tuned its cutoff frequency relative to that of the bent fibre, enabling controlled tuning across the avoided-crossing region. This approach provides a means of manipulating the effective axial potential and the localized resonant states without directly modifying the resonator geometry.

The experimental observations closely reflect the quantum-mechanical physics of avoided crossings between coupled states. In atomic and molecular systems, two diabatic potential-energy curves approach and would cross in the absence of interaction. Coupling between the states lifts this degeneracy and produces repulsive adiabatic potential curves. In our photonic system, the uncoupled cutoff frequencies of the two fibres play the role of diabatic potentials, while their inter-fibre coupling produces the corresponding adiabatic potentials. As the straight-fibre cutoff frequency approaches that of the bent fibre, the coupling-induced potential well deepens and the resonator depth increases, analogous to the enhanced hybridization of quantum states near an avoided crossing. Beyond the crossing, the resonator depth decreases as the frequency detuning increases. Moreover, the experimentally observed tendency of the localized resonant spectrum to remain near the bent-fibre cutoff frequency, while the straight-fibre cutoff frequency continues to increase, is consistent with the exchange of modal character between the adiabatic branches. This behaviour gives rise to the characteristic apparent “hanging” of the resonator spectrum below the shifting cutoff-frequency trace of the stretched fibre. Thus, our experiment provides a directly tunable photonic realization of avoided-crossing physics in which the evolution of coupled states and their effective potential profiles can be controlled mechanically and observed through spectral measurements.

With perturbative corrections from surrounding WGMs, our adiabatic two-mode model fits the measured spectrograms well and yields a stretch-induced cutoff shift that follows the tuning trend predicted by the first-order geometric and elasto-optic model for silica. Generally, other WGMs may have cutoffs close to the selected pair even before stretching begins. When the detuning of one of these modes from either branch is comparable to its coupling to that branch, it can reshape the created microresonator and needs to be included explicitly. An important next step is to calculate the neighbouring cutoff frequencies and coupling strengths and to extract both from spectrograms recorded over a wider frequency range and at finer stretching steps. It would also be interesting to investigate situations when the considered adiabatic approximation breaks down and the full axial coupled wave equations need to be solved, both for the original pair of modes and for a larger set of coupled modes.

These results establish mechanical stress as a practical control parameter for engineering coupling-induced optical potentials and their localized states. In contrast to conventional mechanical tuning, stretching the straight fibre through that of the bent fibre changing how their WGMs hybridise and thereby reconfiguring the created microresonator spectrum. By stretching the straight fibre, we can vary this detuning directly to adjust both the depth and the spectral position of the induced potential well. This adds the previously missing degree of freedom for tuning microresonators formed by side coupling between optical fibres. The developed approach therefore provides a versatile platform for investigating avoided-crossing phenomena in photonic systems. The combination of stretching with the previously demonstrated translation and bending of side-coupled optical fibres enable widely reconfigurable, low loss and precise photonic devices for applications in optical signal processing and sensing.

**Acknowledgement**
This work was supported by the Engineering and Physical Sciences Research Council (EPSRC) (grants EP/X03772X/1 and EP/W002868/1).

**Author contributions**
M.S. conceived the idea and supervised the project; I.S. developed the setup and performed the experiments; M.S. developed the theoretical model. Both authors analysed the data and prepared the manuscript.

**Data availability**
The data that support the findings of this study are available from the corresponding author upon reasonable request.

**Conflict of interest**
The authors declare no conflict of interest.

# Supplementary Information

## S1. Second-order reduction for two nearby modes

At each axial coordinate $z$, the adiabatic potentials $U(z)$ are the eigenvalues of the local Hermitian matrix $\mathbf{V}(z)$. Following Eq. (1) of the main text, the diagonal elements are $V_{nn}(z) = \nu_{c,n}(z) - \nu_0$, where $\nu_{c,n}$ is the uncoupled cutoff frequency and $\nu_0$ is the common reference frequency. The off-diagonal elements $V_{mn}(z) = V_{nm}^*(z)$ describe interchannel coupling. The local eigenvector $\mathbf{c}(z)$ satisfies

$$\mathbf{V}(z)\mathbf{c}(z) = U(z)\mathbf{c}(z), \qquad \sum_{n=1}^{N} V_{mn}(z)c_n(z) = U(z)c_m(z). \tag{S1}$$

The retained subspace $P$ contains channels 1 and 2, for which the detuning is comparable to the direct coupling near the avoided crossing. The complementary subspace $Q$ contains channels $n = 3, \dots, N$. Suppressing $z$ for clarity, the eigenvalue problem becomes

$$\begin{pmatrix} \mathbf{V}_{PP} & \mathbf{V}_{PQ} \\ \mathbf{V}_{QP} & \mathbf{V}_{QQ} \end{pmatrix} \begin{pmatrix} \mathbf{c}_P \\ \mathbf{c}_Q \end{pmatrix} = U \begin{pmatrix} \mathbf{c}_P \\ \mathbf{c}_Q \end{pmatrix}. \tag{S2}$$

Solving the second block row for $\mathbf{c}_Q$ and substituting into the first gives

$$\begin{aligned} \mathbf{c}_Q &= \left(U\mathbf{I}_Q - \mathbf{V}_{QQ}\right)^{-1}\mathbf{V}_{QP}\mathbf{c}_P, \\ \left[\mathbf{V}_{PP} + \mathbf{V}_{PQ}\left(U\mathbf{I}_Q - \mathbf{V}_{QQ}\right)^{-1}\mathbf{V}_{QP}\right]\mathbf{c}_P &= U\mathbf{c}_P, \end{aligned} \tag{S3}$$

where $\mathbf{I}_Q$ is the identity on $Q$. Equation (S3) is exact where the inverse exists. Let $U_\alpha^{(0)}$ and $\mathbf{u}_\alpha$ be the eigenvalues and normalised eigenvectors of $\mathbf{V}_{PP}$, with $\alpha = \pm$. For a perturbative reduction, the appreciably coupled remote modes must be separated from both retained branches. The combined remote admixture must also be small for each branch, as assessed from

$$\mathbf{q}_\alpha = \left(U_\alpha^{(0)}\mathbf{I}_Q - \mathbf{V}_{QQ}\right)^{-1}\mathbf{V}_{QP}\mathbf{u}_\alpha, \qquad \mathbf{q}_\alpha^\dagger\mathbf{q}_\alpha \ll 1, \quad \alpha = \pm. \tag{S4}$$

These conditions apply over the region of appreciable coupling at each elongation used in the fit. To obtain the leading remote amplitudes explicitly, write the second block row of Eq. (S2) for an individual channel $n$. The remote amplitudes vanish at zeroth order, so $c_n^{(0)} = 0$. Counting the off-diagonal couplings involving $Q$ as first order on the scale of the remote detunings gives

$$(U - V_{nn})c_n = \sum_{j=1}^{2} V_{nj}\, c_j + \sum_{\substack{m=3 \\ m\neq n}}^{N} V_{nm}\, c_m,$$
$$c_n^{(1)} = \sum_{j=1}^{2} \frac{V_{nj}c_j}{U - V_{nn}}, \qquad n = 3, \ldots, N. \tag{S5}$$

The second sum in the first line is second order because $c_m$ starts at first order. Substitution of $c_n^{(1)}$ into the first block row produces the energy-dependent second-order correction

$$\sum_{j=1}^{2}\left[V_{ij} + \Sigma_{ij}(U)\right] c_j = U c_i,$$
$$\Sigma_{ij}(U) = \sum_{n=3}^{N} \frac{V_{in}V_{nj}}{U - V_{nn}}, \qquad i,j = 1,2. \tag{S6}$$

For the energy-independent reduction used in the fit, we also take $|V_{12}|$ to be small on the scale of the remote detunings. The diabatic cutoffs can then serve as the unperturbed values when constructing the second-order correction. This ordering allows $|V_{12}|$ to be comparable to the detuning within the retained pair.

Let $\mathbf{D}_P$ and $\mathbf{D}_Q$ denote the diagonal parts of the retained and remote blocks. The leading admixture of remote channel $n$ into retained channel $i$ is $B_{ni} = V_{ni}/(V_{ii} - V_{nn})$. The columns of $\binom{\mathbf{I}_P}{\mathbf{B}}$ have overlap matrix $\mathbf{I}_P + \mathbf{B}^\dagger\mathbf{B}$. Orthonormalising these columns to second order adds the normalisation terms to the projected matrix. Using $\mathbf{B}\mathbf{D}_P - \mathbf{D}_Q\mathbf{B} = \mathbf{V}_{QP}$, the correction reduces to $\left(\mathbf{V}_{PQ}\mathbf{B} + \mathbf{B}^\dagger\mathbf{V}_{QP}\right)/2$. Thus

$$\left(V_{\text{eff}}^{(2)}\right)_{ij} = V_{ij} + \frac{1}{2}\sum_{n=3}^{N} V_{in}\, V_{nj}\left(\frac{1}{V_{ii} - V_{nn}} + \frac{1}{V_{jj} - V_{nn}}\right), \quad i,j = 1,2. \tag{S7}$$

The symmetrised denominators make the matrix Hermitian. We retain the direct coupling $V_{12}$ in the effective matrix and diagonalise the resulting two-mode block exactly. In matrix form,

$$\mathbf{V}_{\text{eff}}^{(2)} = \begin{pmatrix} V_{11} + \Delta_1^{(2)} & V_{12} + K_{12}^{(2)} \\ V_{21} + K_{21}^{(2)} & V_{22} + \Delta_2^{(2)} \end{pmatrix}. \tag{S8}$$

The diagonal corrections are

$$\Delta_j^{(2)}(z) = \sum_{n=3}^{N} \frac{|V_{jn}(z)|^2}{V_{jj}(z) - V_{nn}(z)}, \qquad j = 1,2. \tag{S9}$$

The off-diagonal corrections are

$$K_{12}^{(2)}(z) = \frac{1}{2}\sum_{n=3}^{N} V_{1n}(z)V_{n2}(z)\left(\frac{1}{V_{11}(z) - V_{nn}(z)} + \frac{1}{V_{22}(z) - V_{nn}(z)}\right),$$
$$K_{21}^{(2)}(z) = K_{12}^{(2)*}(z). \tag{S10}$$

In the side-coupled fibre model [S1], the retained modes belong to different fibres. Each remote mode belongs to one uncoupled fibre. The uncoupled modes on each fibre are orthogonal channels, and the off-diagonal coupling connects modes on different fibres. Thus each remote mode couples to at most one of the retained modes and

$$V_{1n}(z)V_{n2}(z) = 0, \qquad K_{12}^{(2)}(z) = K_{21}^{(2)}(z) = 0. \tag{S11}$$

The effective diagonal elements are $V_j^{\text{eff}} = V_{jj} + \Delta_j^{(2)}$. With $\overline{V} = \left(V_1^{\text{eff}} + V_2^{\text{eff}}\right)/2$ and $\Delta V = V_1^{\text{eff}} - V_2^{\text{eff}}$, the adiabatic potentials are

$$U^{\pm}(z) = \overline{V}(z) \pm \sqrt{\frac{\Delta V(z)^2}{4} + |V_{12}(z)|^2}. \tag{S12}$$

Equation (S12) describes the local avoided crossing under the remote-mode reduction. Using these potentials to describe axial propagation also requires sufficiently slow axial variation to suppress conversion between the local branches.

## S2. Strain-induced cutoff shift and actuation force

For a WGM of fixed azimuthal and radial order, the cutoff frequency scales to leading order as $\nu_c \propto (n_{\mathrm{eff}} r_0)^{-1}$ [S2]. A small uniform deformation therefore gives

$$\frac{\Delta\nu_c}{\nu_c} = -\frac{\Delta r_0}{r_0} - \frac{\Delta n_{\mathrm{eff}}}{n_{\mathrm{eff}}}. \tag{S13}$$

Let $s = \Delta L/L$ be the actual tensile strain of the free fibre segment, with unstrained length $L$ and actual elongation $\Delta L$. For a polarisation family $q$, the radial contraction and effective-index change are

$$\frac{\Delta r_0}{r_0} = -\mu_P s, \qquad \frac{\Delta n_{\mathrm{eff},q}}{n_{\mathrm{eff},q}} = -p_{e,q} s, \tag{S14}$$

where $\mu_P$ is the Poisson ratio and $p_{e,q}$ is the effective photoelastic coefficient. Substituting Eq. (S14) into Eq. (S13) gives Eq. (9) of the main text,

$$\frac{\Delta\nu_{c,q}}{\nu_{c,q}} = C_q s = C_q \frac{\Delta L}{L}, \qquad C_q = \mu_P + p_{e,q}. \tag{S15}$$

Both radial contraction and a decrease in refractive index increase the cutoff frequency. For isotropic silica under uniaxial strain, the transverse and axial photoelastic coefficients are [S2, S3]

$$\begin{aligned} p_{e,\perp} &= \frac{n^2}{2}[p_{12} - \mu_P(p_{11} + p_{12})], \\ p_{e,\parallel} &= \frac{n^2}{2}(p_{11} - 2\mu_P p_{12}). \end{aligned} \tag{S16}$$

The values $p_{11} = 0.116$ and $p_{12} = 0.255$ measured at 1531 nm [S2], together with $n = 1.444$ [S4] and $\mu_P = 0.17$ [S3, S5], give $p_{e,\perp} \simeq 0.200$ and $p_{e,\parallel} \simeq 0.031$. For the numerical estimates, we adopt the measured WGM tuning coefficients near 1531 nm in frequency units, $C_{\mathrm{TE}} = 0.369$ and $C_{\mathrm{TM}} = 0.201$ [S2]. At a given frequency shift, the TM estimate of strain, elongation and force is larger by a factor $C_{\mathrm{TE}}/C_{\mathrm{TM}} \simeq 1.84$.

Solving Eq. (S15) for strain and elongation gives

$$s = \frac{\Delta\nu_{c,q}}{C_q \nu_{c,q}}, \qquad \Delta L = \frac{L\,\Delta\nu_{c,q}}{C_q \nu_{c,q}}. \tag{S17}$$

In the experiment, the straight, stretched fibre is fibre 1 and the bent fibre is fibre 2. The length of the stretched segment was $L = 55$ mm. The nominal stage displacement $\Delta L_{st}$ took the values 0, 20, 40, 60, 80 and 100 µm. The fitted cutoff intercept of fibre 1 at $z = 0$ is $a_1(\Delta L_{st})$, as defined in Section S3. The fitted shift and the prediction for full transfer of the nominal displacement to the free fibre are

$$\begin{aligned} \Delta\nu_{c,1}^{\mathrm{fit}}(\Delta L_{st}) &= a_1(\Delta L_{st}) - a_1(0), \\ \Delta\nu_{c,1}^{\mathrm{ideal}}(\Delta L_{st}) &= C_{\mathrm{TE}} a_1(0) \frac{\Delta L_{st}}{L}. \end{aligned} \tag{S18}$$

With $a_1(0) = 193.221915$ THz, the ideal tuning rate is 1.296 GHz/µm, or 25.93 GHz per 20 µm of elongation. The fitted shifts are 0, 25.721, 49.021, 71.661, 91.542 and 111.424 GHz. A fit constrained to pass through the origin gives 1.147 GHz/µm, about 11.5% below the ideal value. Figure S1 compares the fitted shifts with Eq. (S18).

The difference may reflect strain transfer through the mounts or a different strain coefficient for the measured WGM. The increments between successive fitted settings range from about 19.88 to 25.72 GHz. Possible contributions include compliance, adhesive creep or slip, nonuniform strain, displacement calibration and correlations between the fitted cutoff intercept and the parameters of the coupling region.

For a bare silica fibre under uniform axial stress, Hooke's law gives $F = Y\pi r_0^2 s$, where $Y$ is Young's modulus [S3]. Taking $Y = 72$ GPa for fused silica [S5] and using Eq. (S17) yields Eq. (10) of the main text,

$$F = \pi r_0^2 Y \frac{\Delta L}{L} = \frac{\pi r_0^2 Y}{C_q} \frac{\Delta \nu_{c,q}}{\nu_{c,q}}. \tag{S19}$$

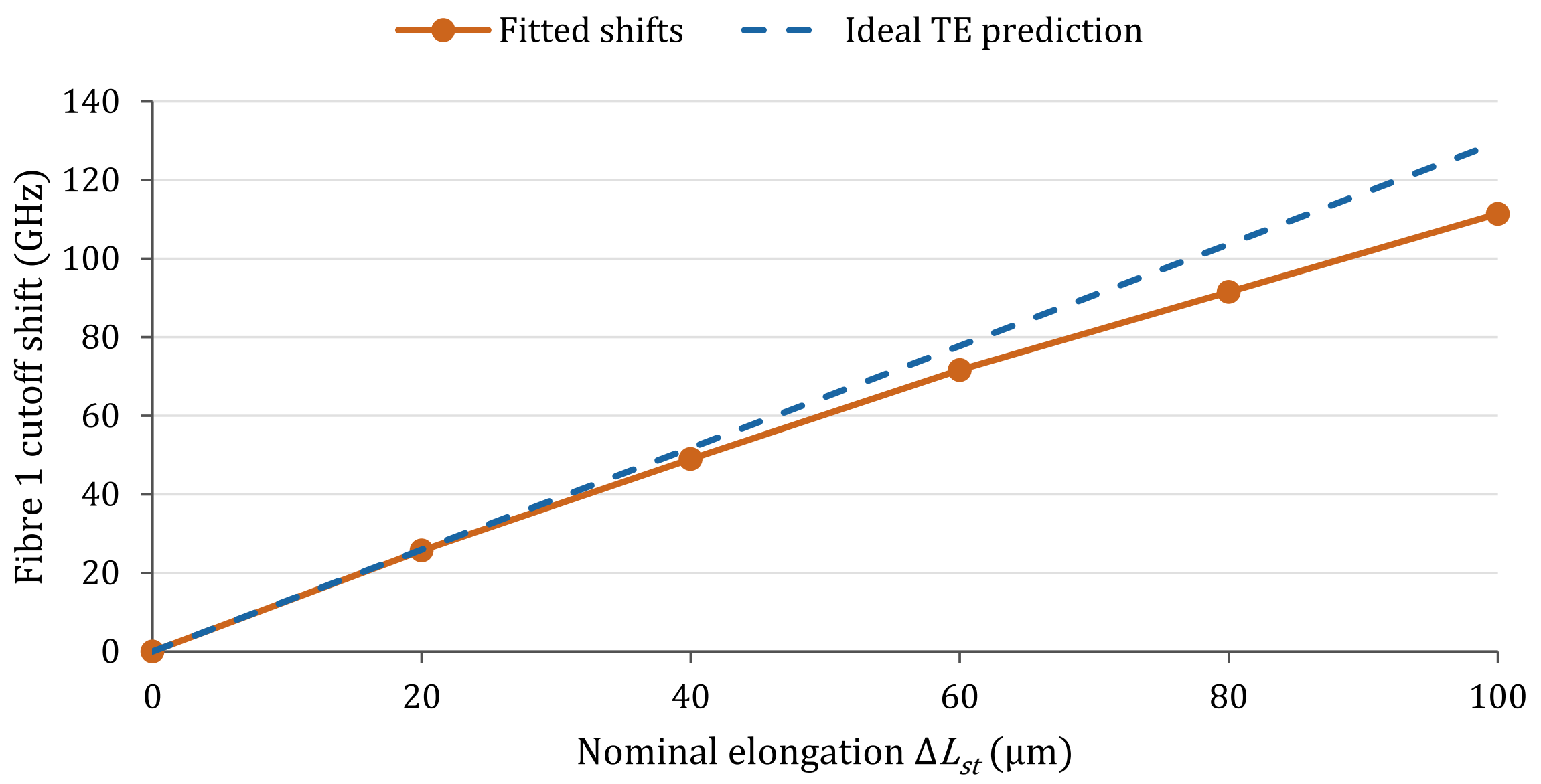

**Figure S1.** Cutoff shift of the straight fibre as a function of its nominal elongation $\Delta L_{st}$. The markers are obtained from $a_1(\Delta L_{st}) - a_1(0)$. The dashed line is the ideal TE prediction from Eq. (S18).

Table S1 uses the experimental radius $r_0 = 40$ μm and length $L = 55$ mm. The force for a fixed cutoff shift scales as $r_0^2$, and the actual elongation for that shift is proportional to $L$.

**Table S1.** Strain, actual elongation and force estimates for a TE-like silica-fibre WGM at $\nu_c = 193$ THz, with $C_{\mathrm{TE}} = 0.369$ and $Y = 72$ GPa.

| **Cutoff shift** | **Strain $s$** | **Actual elongation** | **Force at $r_0 = 40$ μm** | **Force at $r_0 = 20$ μm** |
|---|---|---|---|---|
| 0.1 GHz | $1.40 \times 10^{-6}$ | 77.2 nm | 0.508 mN | 0.127 mN |
| 1 GHz | $1.40 \times 10^{-5}$ | 0.772 μm | 5.08 mN | 1.27 mN |
| 10 GHz | $1.40 \times 10^{-4}$ | 7.72 μm | 50.8 mN | 12.7 mN |
| 100 GHz | $1.40 \times 10^{-3}$ | 77.2 μm | 0.508 N | 0.127 N |

For the experimental fibre, a 100 MHz shift corresponds to an actual elongation of 77.2 nm and a force of 0.508 mN. A 1 GHz shift requires 5.08 mN. For comparison, an electrothermal stage achieved an 85 μm range of motion and a maximum force of 11.8 mN [S6], and an electrostatic MEMS gripper reported a gripping force of 15 mN over its 1 mm range [S7]. A 100 GHz shift in the experimental fibre requires approximately 0.51 N, which sets the force scale for actuation over the full tuning range.

## S3. Fitting the cutoff frequency profiles

The six coupled-fibre spectrograms in Figs. 4 and 6 of the main text correspond to the six nominal elongations $\Delta L_{st}$ specified in Section S2. The separate bent-fibre spectrogram provides the uncoupled reference for fibre 2. The uncoupled physical cutoff frequencies at each elongation are modelled as

$$\nu_{c,j}(z,\Delta L_{st}) = a_j(\Delta L_{st}) + b_j z, \qquad j = 1,2. \tag{S20}$$

Here $a_j(\Delta L_{st})$ is the physical cutoff intercept at $z = 0$, and $b_j$ is the axial slope, common to all elongations for fibre $j$. The corresponding diabatic potentials are measured relative to the same reference frequency $\nu_0$ as in the main text,

$$V_{jj}(z,\Delta L_{st}) = \nu_{c,j}(z,\Delta L_{st}) - \nu_0, \qquad j = 1,2. \tag{S21}$$

We use the Gaussian envelope of main-text Eq. (14),

$$G(\zeta) = \exp\left(-\frac{\zeta^2}{z_0^2}\right), \qquad \zeta = z - z_c(\Delta L_{st}). \tag{S22}$$

The interaction centre $z_c(\Delta L_{st})$ can shift with elongation. The width parameter $z_0$ and the peak direct coupling $W_0$ are common to all elongations. The semiclassical approximation gives a common axial dependence for couplings between nearby cutoffs [S8]. We therefore use the same Gaussian envelope for the retained pair and the couplings to remote modes,

$$\begin{aligned} V_{12}(z,\Delta L_{st}) &= W_0 G(\zeta), \\ V_{jn}(z,\Delta L_{st}) &\simeq w_{jn}(\Delta L_{st}) G(\zeta), \qquad j = 1,2, \quad n = 3, \dots, N. \end{aligned} \tag{S23}$$

The relative phases of the retained modes are chosen so that $W_0$ is real and nonnegative. Here $w_{jn}(\Delta L_{st}) = V_{jn}(z_c,\Delta L_{st})$, with $z_c = z_c(\Delta L_{st})$ for brevity. Define the signed detuning $D_{jn} = V_{jj} - V_{nn}$. If this detuning varies by a small fraction over the interaction region, Eq. (S9) gives the centre-value approximation

$$\begin{aligned} \Delta_j^{(2)}(z,\Delta L_{st}) &\simeq A_j(\Delta L_{st}) G^2(\zeta), \\ A_j(\Delta L_{st}) &= \sum_{n=3}^{N} \frac{|w_{jn}(\Delta L_{st})|^2}{D_{jn}(z_c,\Delta L_{st})}, \qquad j = 1,2. \end{aligned} \tag{S24}$$

The common envelope accounts for the $G^2$ dependence of the diagonal corrections. The peak overlaps can differ between modes. The fit treats $A_j(\Delta L_{st})$ as free amplitudes at the measured elongations.

The corrected diagonal elements in physical frequency units are

$$\tilde{\nu}_j(z,\Delta L_{st}) = a_j(\Delta L_{st}) + b_j z + A_j(\Delta L_{st}) G^2(\zeta), \qquad j = 1,2. \tag{S25}$$

The effective potentials are $V_j^{\text{eff}} = \tilde{\nu}_j - \nu_0$. With $\bar{\nu} = (\tilde{\nu}_1 + \tilde{\nu}_2)/2$ and $\Delta\nu = \tilde{\nu}_1 - \tilde{\nu}_2$, Eq. (S12) gives the two fitted local cutoff profiles,

$$\begin{aligned} \nu_c^{\pm}(z,\Delta L_{st}) &= \bar{\nu} \pm \sqrt{\frac{\Delta\nu^2}{4} + W_0^2 G^2(\zeta)}, \\ U^{\pm}(z,\Delta L_{st}) &= \nu_c^{\pm}(z,\Delta L_{st}) - \nu_0. \end{aligned} \tag{S26}$$

Equation (S26) diagonalises the direct coupling within the retained pair. The curves compared with the measured spectrograms in Fig. 6 of the main text are $\nu_c^{\pm}$ in physical frequency units.

**Table S2.** Definitions and units of the fitted model quantities.

| Quantity | Definition | Scope | Units |
|---|---|---|---|
| $\Delta L_{st}$ | Nominal elongation of fibre 1 | Experimental setting | µm |
| $a_j(\Delta L_{st})$ | Uncoupled cutoff frequency of fibre $j$ at $z = 0$ | Elongation dependent | THz |
| $b_j$ | Axial slope of the uncoupled cutoff frequency | Common for each fibre | GHz/mm |
| $z_c(\Delta L_{st})$ | Axial centre of the interaction | Elongation dependent | mm |
| $z_0$ | Gaussian width parameter | Common | mm |
| $W_0$ | Peak direct coupling of the retained modes | Common | GHz |
| $A_j(\Delta L_{st})$ | Peak diagonal correction for fibre $j$ | Elongation dependent | GHz |
| $G(\zeta)$ | Dimensionless Gaussian coupling envelope | Derived function | — |
| $U^{\pm}(z,\Delta L_{st})$ | Local adiabatic potentials relative to $\nu_0$ | Model output | THz |
| $\nu_c^{\pm}(z,\Delta L_{st})$ | Physical frequencies of the local coupled cutoffs | Model output | THz |

The common values of $b_1$, $b_2$, $W_0$ and $z_0$ are listed in Table S3. The full width at half maximum of $G$ is $2\sqrt{\ln 2}\, z_0 = 0.3223$ mm.

**Table S3.** Common parameters used at all nominal elongations.

| Quantity | Value | Units |
|---|---|---|
| $b_1$ | 26.980785 | GHz/mm |
| $b_2$ | 16.900000 | GHz/mm |
| $W_0$ | 41.583740 | GHz |
| $z_0$ | 0.193534618 | mm |

Table S4 lists the parameters that depend on nominal elongation. The intercepts and diagonal corrections are paired by fibre. Convert $A_j$ from GHz to THz when using the values in Eq. (S25) with the intercepts in THz.

**Table S4.** Elongation-dependent fit parameters.

| $\Delta L_{st}$ (µm) | $a_1(\Delta L_{st})$ (THz) | $a_2(\Delta L_{st})$ (THz) | $z_c(\Delta L_{st})$ (mm) | $A_1(\Delta L_{st})$ (GHz) | $A_2(\Delta L_{st})$ (GHz) |
|---|---|---|---|---|---|
| 0 | 193.221915 | 193.278631 | 0.829465 | −32.937 | −33.942 |
| 20 | 193.247636 | 193.281761 | 0.778001 | +2.133 | −61.829 |
| 40 | 193.270936 | 193.284842 | 0.779137 | −43.708 | −15.084 |
| 60 | 193.293576 | 193.285337 | 0.771816 | −55.155 | −4.005 |
| 80 | 193.313457 | 193.286494 | 0.742420 | −68.506 | +6.477 |
| 100 | 193.333339 | 193.287863 | 0.701707 | −37.813 | −28.621 |

Figure S2 shows the fitted cutoff intercepts. The intercept of the stretched fibre increases by approximately 111.4 GHz over the measured range, while the intercept of the bent fibre varies by about 9.2 GHz.

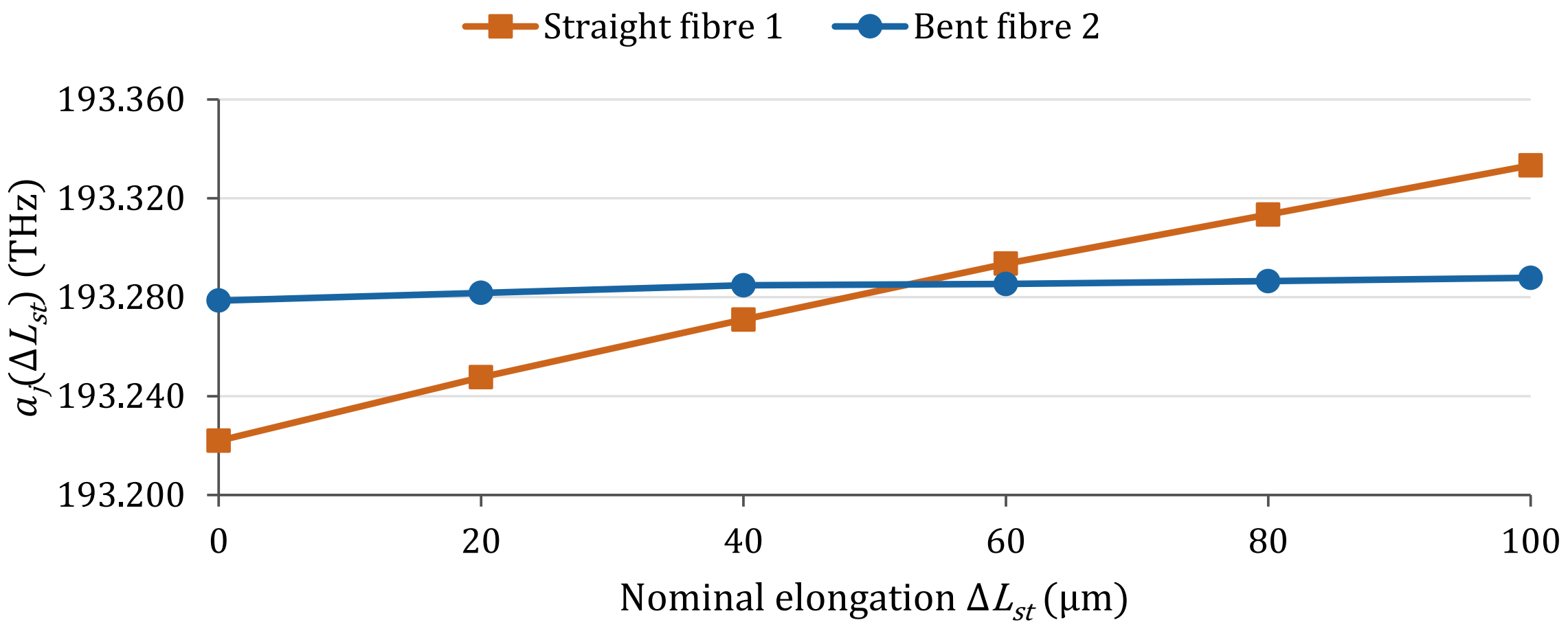

**Figure S2.** Cutoff intercepts $a_1(\Delta L_{st})$ and $a_2(\Delta L_{st})$ for the straight and bent fibres, respectively.

Figure S3 shows the fitted interaction centre, which shifts from about 0.829 mm to 0.702 mm across the measured range.

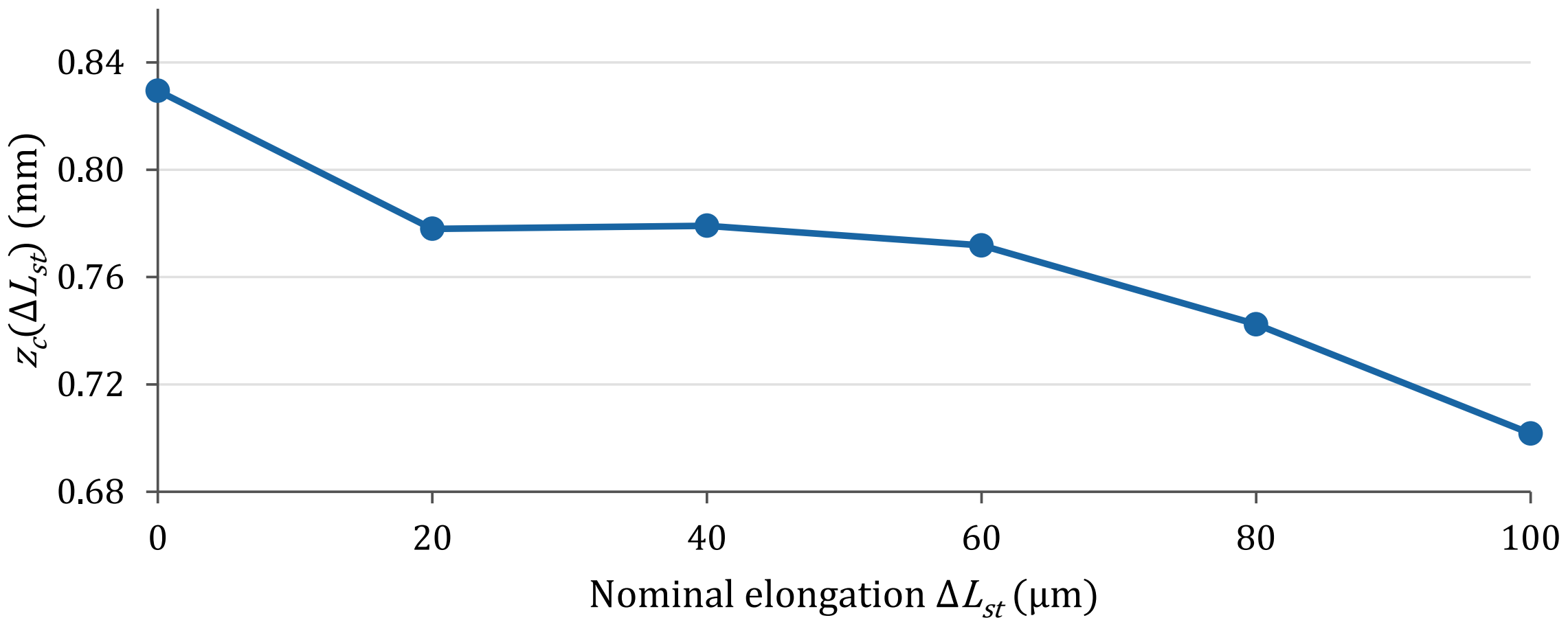

**Figure S3.** Fitted interaction centre $z_c(\Delta L_{st})$ as a function of nominal elongation.

Figure S4 shows that, from 20 to 80 μm, $A_1$ decreases and $A_2$ increases, with their sum ranging from −62.0 to −58.8 GHz. Equation (S24) gives these directions of change if the signed detunings increase for fibre 1 and decrease for fibre 2, the overlaps are approximately fixed, and the denominators stay away from zero. Remote-mode crossings between measured elongations could contribute to the nonmonotonicity over the full sweep. At each measured setting, the reduction requires sufficiently small remote admixture across the coupling region, as assessed by Eq. (S4). A mode that violates this condition must be included explicitly.

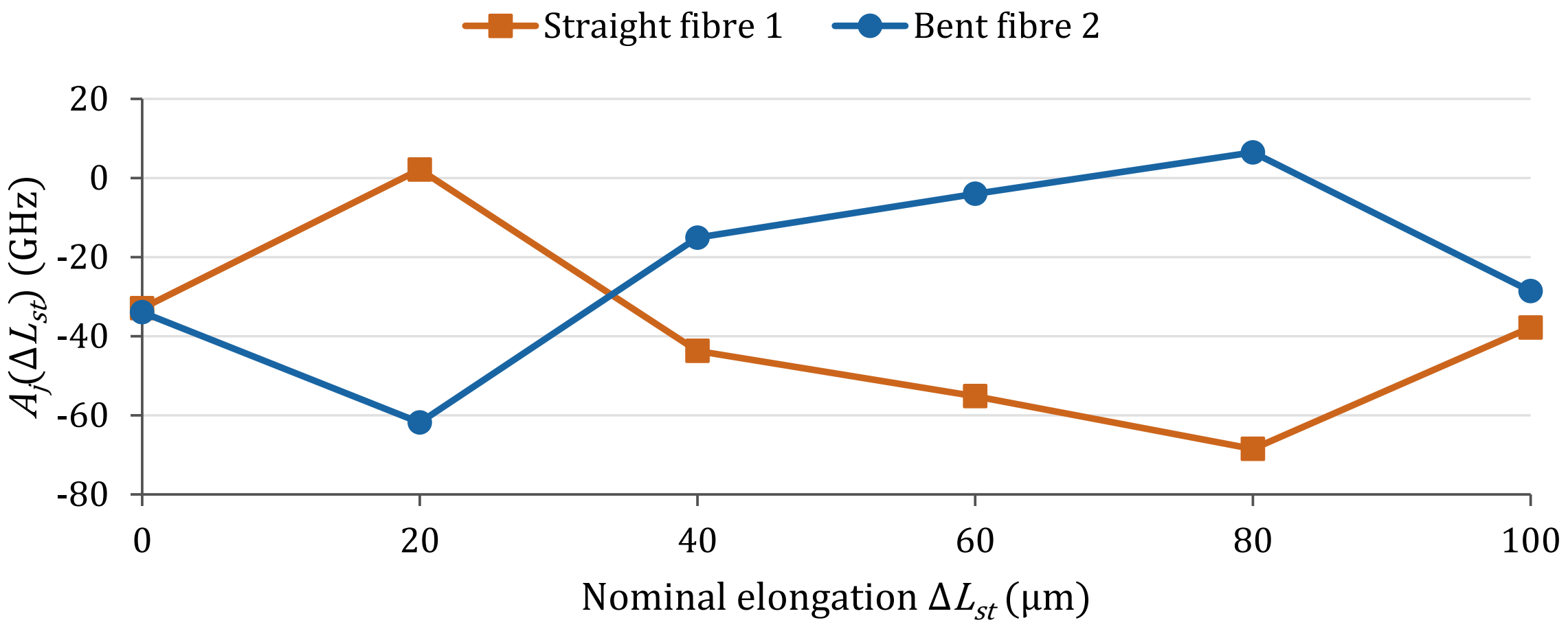

**Figure S4.** Fitted second-order diagonal corrections $A_1(\Delta L_{st})$ and $A_2(\Delta L_{st})$ for the straight and bent fibres, respectively.

## Supplementary references